\documentclass{iopart}
\usepackage{graphicx}  
\usepackage{latexsym}

\jl{6}        
\eqnobysec    

\def\beq{\begin{equation}}
\def\eeq{\end{equation}}

\def\rmd{{\rm d}}

\begin{document}

\title[Emission vs Fermi coordinates]
{Emission vs Fermi coordinates: applications to relativistic positioning systems}

\author{
D Bini$^* {}^\S{}^\P$, A Geralico$^\S{}^\dag$, M L Ruggiero$^\ddag{}^{||}$ and A Tartaglia$^\ddag{}^{||}$}
\address{
  ${}^*$\
Istituto per le Applicazioni del Calcolo ``M. Picone,'' CNR I-00161 Rome, Italy
}
\address{
  ${}^\S$\
  ICRA, University of Rome ``La Sapienza,'' I-00185 Rome, Italy
  }
\address{
${}^\P$\
  INFN - Sezione di Firenze, Polo Scientifico, Via Sansone 1, 
  I-50019, Sesto Fiorentino (FI), Italy 
}
\address{
  ${}^\dag$\
  Physics Department, University of Rome ``La Sapienza,'' I-00185 Rome, Italy
}
\address{
$^\ddag$\
Dipartimento di Fisica, Politecnico di Torino, Corso Duca degli Abruzzi 24, I-10129 Torino, Italy
}
\address{
${}^{||}$\
  INFN - Sezione di Torino, Via P. Giuria 1, 
  I-10125, Torino, Italy 
}

\begin{abstract}
A 4-dimensional relativistic positioning system for a general spacetime is constructed by using the so called \lq\lq emission coordinates." 
The results apply in a small region around the world line of an accelerated observer carrying a Fermi triad, as described by the Fermi metric. 
In the case of a Schwarzschild spacetime modeling the gravitational field around the Earth and 
an observer at rest at a fixed spacetime point, these coordinates realize a relativistic positioning system alternative to the current GPS system. 
The latter is indeed essentially conceived as Newtonian, so that it necessarily needs taking into account at least the most important relativistic effects through Post-Newtonian corrections to work properly.
Previous results concerning emission coordinates in flat spacetime are thus extended to this more general situation.
Furthermore, the mapping between spacetime coordinates and emission coordinates is completely determined by means of the world function, which in the case of a Fermi metric can be explicitly obtained.
\end{abstract}

\pacno{04.20.Cv}

\section{Introduction}

Currently there is a growing interest in the construction of an emission coordinate system for the Earth in order to improve
the current positioning systems (GPS, GLONASS) \cite{ashby,sanchez}. In fact, the latter are essentially conceived as Newtonian, hence based on a classical (Euclidean) space and absolute time, over which some relativistic corrections are added via the Post-Newtonian formalism.
On the contrary, emission coordinates allow the definition of {\it relativistic} positioning systems whose study  based on the framework and the concepts of general relativity was initiated by Coll and collaborators several years ago \cite{coll1,coll2,coll3,coll4,coll5,coll6}. 

At present global positioning systems consist of a non-inertial spatial reference frame for navigation that co-rotates with the Earth and is geocentric (the ECEF, Earth Centered Earth Fixed system), and on the coordinate time of a local inertial \lq\lq star-fixed'' reference frame whose origin lies at the Earth's center of mass and which is freely falling with it (the ECI, Earth Centered Inertial system).
Although clock speeds are small in comparison with the speed of light and gravitational fields are weak near the Earth, they give rise
to significant relativistic effects. 
The most important ones affecting the rate of clocks (to the order $1/c^2$, which is just the order of approximation
used in GPS) are first and second order Doppler frequency shifts of clocks due to their relative motion, gravitational frequency shifts due to the Earth's mass, and the Sagnac effect
due to the Earth's rotation \cite{ashby,sanchez,bahder}. 
If these corrections were not made, the system would not be operational after a few minutes. In a day of operation, they would produce an error of more than 11 km in the horizontal positioning of the receiver. In a week, the error accumulated in the vertical positioning would be approximately  5 km. 
Therefore, fully-relativistic positioning systems are not only conceptually simpler than GPS systems, but also more accurate, because no corrections are needed at all, whereas Post-Newtonian corrections to current positioning systems are included up to the order $1/c^2$ only.

Recently the project SYPOR \cite{coll1} (SYst\`eme de POsitionnement Relativiste) has proposed to endow the constellation of satellites of GALILEO with the necessary elements to constitute by itself a primary, autonomous positioning system for the Earth and its neighbors, referring to the capability of the constellation
to provide complete relativistic metric information, i.e., to describe both the kinematics and the dynamics of the constellation itself as well as of the users. 
In this primary positioning system, an observer at any event in a given spacetime region can know its proper
coordinates.
The system is also autonomous or autolocated if any receiver
determines its spacetime path as well as the trajectories of the satellites solely on the basis of the
information received during a proper time interval.
Hence, a constellation of satellites with clocks that interchange their proper time among themselves and with Earth receivers is a fully relativistic system.
According to the SYPOR project the GALILEO system would not need \lq\lq relativistic'' corrections.
Giving a theoretical contribution to that project is  the main  motivation of the present work.

The simplest relativistic positioning system is the one formed by electromagnetic signals broadcasting the proper times $\tau^A$ ($A=1,\ldots,4$) of four independent clocks carried by satellites \footnote{Hereafter we will always use the single word \lq\lq satellite" to mean \lq\lq clock carried by satellite,"  for convenience.} which move along geodesic (i.e., freely falling) world lines.
The above signals, parameterized by the proper time of the clocks, realize four emission coordinates $\{\tau^A\}$.
More precisely, let us consider an arbitrary spacetime covered by the coordinate patch $\{x^\alpha\}$.
Let $\bar P$ be a generic spacetime event with coordinates ${\bar X}^\alpha$ and let $P_A$ be a generic point with coordinates $X_A^\alpha$ along the world line of the $A$th satellite.
The condition ensuring that the signals emitted by the four satellites at $P_A$ meet the receiver location at $\bar P$ is given by
\beq
\label{deflightcone}
\Omega(X_A,{\bar X})=0\ , \qquad X_A^0<{\bar X}^0\ ,
\eeq
where $\Omega(X_A,{\bar X})$ is the world function connecting the receiver and emission points and ${\bar P}$ is in the strict causal future of $P_A$. 
For any given background metric, the world function  is defined as half the square of the spacetime distance between two generic points $x_A$ and $x_B$ connected by a geodesic path (see Eq. (1) of Chap. 2 in Ref. \cite{synge})
\beq
\label{wf_def}
\Omega(x_A,x_B)=\frac12\int_0^1g_{\mu\nu}(x^\alpha)\bigg|_{x^\alpha=x^\alpha(\lambda)}\frac{\rmd x^\mu}{\rmd\lambda}\frac{\rmd x^\nu}{\rmd\lambda}\rmd\lambda\ ,
\eeq
where $x^\alpha(\lambda)$ satisfies the geodesic equation and the affine parameter is such that $x^\alpha(0)=x^\alpha_A$ and $x^\alpha(1)=x^\alpha_B$. 
Eq. (\ref{deflightcone}) is a system of four equations which must be solved for the unknown coordinates ${\bar X}^\alpha$ of $\bar P$ in terms of the satellite coordinates $X_A^\alpha$, supposed to be known in terms of the proper times $\tau^A$ of the satellites, i.e.,  
the emission coordinates of the point $\bar P$.
The calculation of the world function in a generic spacetime is not a trivial task. It is generally performed perturbatively, unless the solution of the geodesic equations is explicitly known, which is usually not the case.

The problem of setting up such an emission coordinate system in the case of flat spacetime has been addressed by Coll and collaborators \cite{coll4,coll5,coll6}, with special interest in the 2-dimensional case, which is the simplest situation to deal with. 
The 4-dimensional case has been considered by Rovelli \cite{rovelli}, who has outlined a procedure to construct a system of emission coordinates (introduced there with the name GPS coordinates) for a particular linear configuration of emitters in flat spacetime, consisting of four satellites moving away from the origin in different directions but at a common speed. 
The geometrical interpretation of Rovelli's construction has been discussed by Blagojevic et al. \cite{hehl}.

In the present paper we explicitly construct emission coordinates 
for a general spacetime, in a small region around the world line of an accelerated observer carrying a Fermi triad, as described by the Fermi metric \cite{mtw}. 
In particular, we study the case of the Schwarzschild spacetime modeling the gravitational field around the Earth and an observer at rest at a fixed point.
  
The procedure is first outlined in flat spacetime, with a convenient choice of satellite motion, leading to simple explicit expressions for the metric components in terms of the new coordinates.
This analysis is then repeated for the more interesting case of the Fermi metric.  Emission coordinates as well as the components of the transformed metric are obtained as corrections to the flat spacetime ones. 
Since the constructing procedure of emission coordinates is completely general, the calculations can be easily extended to different choices of satellite motion.

\section{Flat spacetime}

Let us briefly review the standard construction of GPS coordinates in flat spacetime \cite{rovelli}, whose generalization to the case of Fermi background metric will be discussed in the next section\footnote{
Note that the signature conventions adopted here are different from those of \cite{rovelli}.}.

Consider Minkowski spacetime in standard Cartesian coordinates $(t,x,y,z)$. 
The corresponding line element is given by
\beq
\rmd s^2=\eta_{\alpha\beta}\rmd x^\alpha\rmd x^\beta=-\rmd t^2+\rmd x^2+\rmd y^2+\rmd z^2\ . 
\eeq 
Let the four satellites be represented by test particles in geodesic motion. 
With this choice of coordinates timelike geodesics are straight lines
\beq
\label{satflat}
x_A^\alpha(\tau^A)\equiv S^\alpha_A=U^\alpha_A\tau^A+S_{0A}^\alpha\ , \qquad A=1,\ldots,4\ 
\eeq
where 
\beq
\label{UAdef}
U_A=\gamma_A[\partial_t+v_An_A^i\partial_i]=\cosh\alpha_A\partial_t+\sinh\alpha_An_A^i\partial_i\ 
\eeq 
are their (constant) 4-velocities and $\tau^A$ is the proper time parametrization along each world line.
In Eq. (\ref{UAdef}) $\gamma_A$ is the Lorentz factor and the linear velocities $v_A$ are related to the rapidity parameters $\alpha_A$ by $v_A=\tanh\alpha_A$; $n_A$ denote the spacelike unit vectors along the spatial directions of motion.
Without any loss of generality, we assume that the satellites all start moving from the origin of the coordinate system $O$; so hereafter we set $S_{0A}^\alpha\equiv0$, and hence 
\beq
\label{satflat2}
S^\alpha_A=U^\alpha_A\tau^A\ .
\eeq

Let us consider now a generic spacetime point $\bar P$ with coordinates ${\bar W}^\alpha$ and the generic point $P_A$ with coordinates $S_A^\alpha$ along the world line of the $A$th satellite corresponding to an elapsed amount of proper time $\tau^A$. 
A photon emitted at $P_A$ follows a null geodesic, i.e., the straight line
\beq
\label{fotflat}
x^\alpha(\lambda)\equiv W^\alpha=K^\alpha\lambda+S_A^\alpha\ ,  
\eeq
where $\lambda$ is an affine parameter.
Such a photon will reach $\bar P$ at a certain value $\bar \lambda$ according to 
\beq
\bar W^\alpha=K^\alpha\bar \lambda+S^\alpha_A\ ,  
\eeq
implying that 
\beq
U^\alpha_A\tau^A-\bar W^\alpha=-K^\alpha\bar \lambda\ .  
\eeq
Taking the norm of both sides we get
\beq
\label{eqtauA}
-(\tau^A)^2+||\bar W||^2-2\tau^A(U_A\cdot{\bar W})=0\ ,
\eeq
since $K$ is a null vector.\\ 
\noindent{\it $\bullet$ Emission vs spacetime coordinates}\\
Solving for $\tau^A$ and selecting the solution corresponding to the past light cone leads to (see Eq. (15) of Ref. \cite{rovelli})
\beq
\label{soltauA}
\tau^A=-(U_A\cdot{\bar W})-\sqrt{(U_A\cdot{\bar W})^2+||\bar W||^2}\ .
\eeq
These equations give the four proper times $\tau^A$ associated with each satellite in terms of the Cartesian coordinates of the generic point $\bar P$ in the spacetime, i.e., $\tau^A=\tau^A(\bar W^0,\ldots, \bar W^3)$. The construction of emission coordinates is briefly sketched in Fig. 1. 

Using Eq. (\ref{soltauA}), one can evaluate the inverse of the transformed metric 
\beq
\label{gABinvflat}
g^{AB}=\eta^{\alpha\beta}\frac{\partial \tau^A}{\partial {\bar W}^\alpha}\frac{\partial \tau^B}{\partial {\bar W}^\beta}\equiv
\eta^{\alpha\beta}(\rmd \tau^A){}_\alpha (\rmd \tau^B){}_\beta=  
\rmd \tau^A \cdot \rmd \tau^B\ ,
\eeq
where the dual frame $(\rmd \tau^A)_\alpha=\partial \tau^A/\partial {\bar W}^\alpha$ also satisfies the following properties
\beq
(\rmd \tau^A){}_\alpha{\bar W}^\alpha=\tau^A\ , \qquad
(\rmd \tau^A){}_\alpha U_A^\alpha=1\ .
\eeq
Similarly one can introduce the frame vectors
\beq
\label{frame}
\left(\frac{\partial}{\partial \tau^A}\right)^\alpha=\frac{\partial \bar W^\alpha}{\partial \tau^A}\ , \qquad (\rmd \tau^A)_\alpha \left(\frac{\partial}{\partial \tau^B}\right)^\alpha=\delta_B^A\ .
\eeq
It is then easy to show \cite{rovelli,hehl} that the condition $g^{AA}=\rmd \tau^A \cdot \rmd \tau^A=0$ is fulfilled.
In fact, by differentiating both sides of Eq. (\ref{eqtauA}) with respect to ${\bar W}^\alpha$ one obtains 
\beq
(\rmd \tau^A)_\alpha=\frac{{\bar W}_\alpha-\tau^AU_A{}_\alpha}{\tau^A+(U_A\cdot {\bar W})}\ ,
\eeq  
which implies 
\beq
g^{AA}=(\rmd \tau^A)^\alpha(\rmd \tau^A)_\alpha=\frac{-(\tau^A)^2+||\bar W||^2-2\tau^A(U_A\cdot{\bar W})}{[\tau^A+(U_A\cdot {\bar W})]^2}=0\ .
\eeq

The metric coefficients $g_{AB}=(\partial / \partial \tau^A) \cdot (\partial / \partial \tau^B)=\eta^{\alpha\beta}(\partial / \partial \tau^A)_\alpha (\partial / \partial \tau^B)_{\beta}$ can be easily obtained as well by expressing the Cartesian coordinates of $\bar P$ in terms of the emission coordinates $\tau^A$, i.e., $\bar W^\alpha = \bar W^\alpha (\tau^1,\ldots, \tau^4)$.\\ 
\noindent{\it $\bullet$ Spacetime vs emission coordinates}\\ 
To accomplish this, it is enough to invert the transformation (\ref{soltauA}).
However, in order to outline a general procedure, we start by considering the equation of the past light cone of the generic spacetime point $\bar P$ with coordinates ${\bar W}^\alpha$ given in terms of the world function, which in the case of flat spacetime is simply given by 
\beq
\label{WFflat}
\Omega_{\rm flat}(x_A,x_B)=\frac12\eta_{\alpha\beta}(x^\alpha_A-x^\alpha_B)(x^\beta_A-x^\beta_B)\ .
\eeq   
The condition (\ref{deflightcone}) ensuring that the past light cone of $\bar P$ cuts the emitter world lines is given by 
\beq
\label{lightcone}
\Omega_{\rm flat}(S_A,{\bar W})=0\ , \qquad S^0_A<{\bar W}^0\ , 
\eeq 
for each satellite labeled by the index $A$. This gives rise to a system of four quadratic equations in the four unknown coordinates ${\bar W}^\alpha$ of the event $\bar P$ of the form (\ref{eqtauA}) for each $A=1,\ldots,4$.
To solve this system start for example by subtracting the last equation from the first three equations to obtain the following system
\begin{eqnarray}\fl\qquad
\label{sistemaflat}
&& \Omega_{\rm flat}(S_i,{\bar W})-\Omega_{\rm flat}(S_4,{\bar W})=0=-2{\bar W}\cdot(S_i-S_4)-(\tau^i)^2+(\tau^4)^2\ ,
\quad i=1,2,3 \nonumber\\
\fl\qquad
&& \Omega_{\rm flat}(S_4,{\bar W})=0=||{\bar W}||^2-2{\bar W}\cdot S_4-(\tau^4)^2\ 
\end{eqnarray}
consisting of three linear equations and only one quadratic equation.
Thus we can  first solve the linear equations for the coordinates ${\bar W}^1, {\bar W}^2, {\bar W}^3$ in terms of ${\bar W}^0$, which then can be determined by the last quadratic equation. As a result, the coordinates of the event $\bar P$ are fully determined in terms of the satellite proper times $\tau^A$ and the known parameters characterizing their world lines.

Consider an example in which one satellite is at rest at the origin $O$ and the other three move along the three spatial axes. Then the $4$-velocities are
\begin{eqnarray}
\label{UAflat}
U_1&=&\cosh \alpha_1\partial_t+\sinh \alpha_1\partial_x\ , \nonumber\\
U_2&=&\cosh \alpha_2\partial_t+\sinh \alpha_2\partial_y\ , \nonumber\\
U_3&=&\cosh \alpha_3\partial_t+\sinh \alpha_3\partial_z\ , \nonumber\\
U_4&=&\partial_t\ , 
\end{eqnarray}
where $\alpha_i$,  $i=1,2,3$, are the rapidities.
The system (\ref{sistemaflat}) then reduces to
\begin{eqnarray}
\label{sistemaexpl}
0&=&\Lambda^i\bar W^0-\bar W^i+\Phi^i\ , \qquad i=1,2,3 \nonumber\\
0&=&-(\bar W^0-\tau^4)^2+\delta_{ij}\bar W^i\bar W^j\ ,
\end{eqnarray}
where the notation 
\begin{eqnarray}
\label{lambda_i}
\Lambda^i=\coth \alpha_i-\frac{\tau^4}{\tau^i \sinh\alpha_i}\ , \qquad
\Phi^i=\frac{(\tau^4)^2-(\tau^i)^2}{2\tau^i\sinh\alpha_i}\ 
\end{eqnarray}
has been introduced.
The solution of Eq. (\ref{sistemaexpl}) is straightforward
\begin{eqnarray}
\label{solsistemaexpl}
{\bar W}^i= \Lambda^i {\bar W}^0+\Phi^i\ , 
\end{eqnarray}
while ${\bar W}^0$ satisfies the quadratic equation 
\beq 
\label{solbarX0flat}
a({\bar W}^0)^2+b{\bar W}^0+c=0 
\eeq 
with coefficients
\begin{eqnarray}\fl\quad
\label{coeffsX0}
a=1-\delta_{ij} \Lambda^i\Lambda^j\ , \quad 
b=-2\left(\tau^4+\delta_{ij}\Lambda^i \Phi^j \right)\ , \quad 
c=(\tau^4)^2-\delta_{ij}\Phi^i\Phi^j\ ,
\end{eqnarray}
which we will assume to be all nonzero hereafter, plus the additional conditions $S^0_A<{\bar W}^0$ ensuring that ${\bar P}$ is in the strict causal future of $P_A$, as stated above.

The components of the frame vectors (\ref{frame}) turn out to be given by
\begin{eqnarray}
\label{xiAdef}
\left(\frac{\partial }{\partial \tau^A}  \right)_0&\equiv&\xi_A=\frac{a}{2a{\bar W}^0+b}\left[\frac{\partial}{\partial{\tau^A}}\left(\frac{c}{a}\right)-{\bar W}^0\frac{\partial}{\partial{\tau^A}}\left(\frac{b}{a}\right)\right]\ , \nonumber \\
\left(\frac{\partial }{\partial \tau^A}  \right)_i&\equiv&\delta_{ij}[Q^j_A+\Lambda^j\xi_A]\ , \qquad Q^i_A=\frac{\partial \Lambda^i}{\partial\tau^A}{\bar W}^0+\frac{\partial \Phi^i}{\partial\tau^A}\ ,
\end{eqnarray}
so that the components $g_{AB}$ of the transformed metric follow easily 
\beq
\label{gABflat}
g_{AB}=-a\xi_A\xi_B+\delta_{ij}[Q^i_AQ^j_B+Q_A^i\Lambda^j\xi_B+\Lambda^i\xi_AQ_B^j]\ .
\eeq
From Eqs. (\ref{xiAdef}) we see that the quantities $\xi_A$ are fractional linear functions of ${\bar W}^0$ whereas the $Q^i_A$ are simply linear functions of ${\bar W}^0$.
As a consequence, each of the metric coefficients can be cast in the form of a fractional linear function of ${\bar W}^0$, i.e.,
\beq
g_{AB}=\frac{a_{AB} {\bar W}^0 +b_{AB}}{c_{AB} {\bar W}^0 +d_{AB}}\ ,
\eeq
where the coefficients $a_{AB}$, $b_{AB}$, $c_{AB}$ and $d_{AB}$ are four independent functions of the emission coordinates $\tau^A$ and of the kinematical parameters of the satellites.
Note that it is easy to show that this is true in general, not only for our particular choice (\ref{UAflat}) of satellite motion.

The same approach we have outlined above will be applied in the next section to the more physically interesting case of a metric describing the homogeneous gravitational field of the Earth.

\begin{figure} 
\typeout{*** EPS figure 1}
\begin{center}
$\begin{array}{ccc}
\includegraphics[scale=0.35]{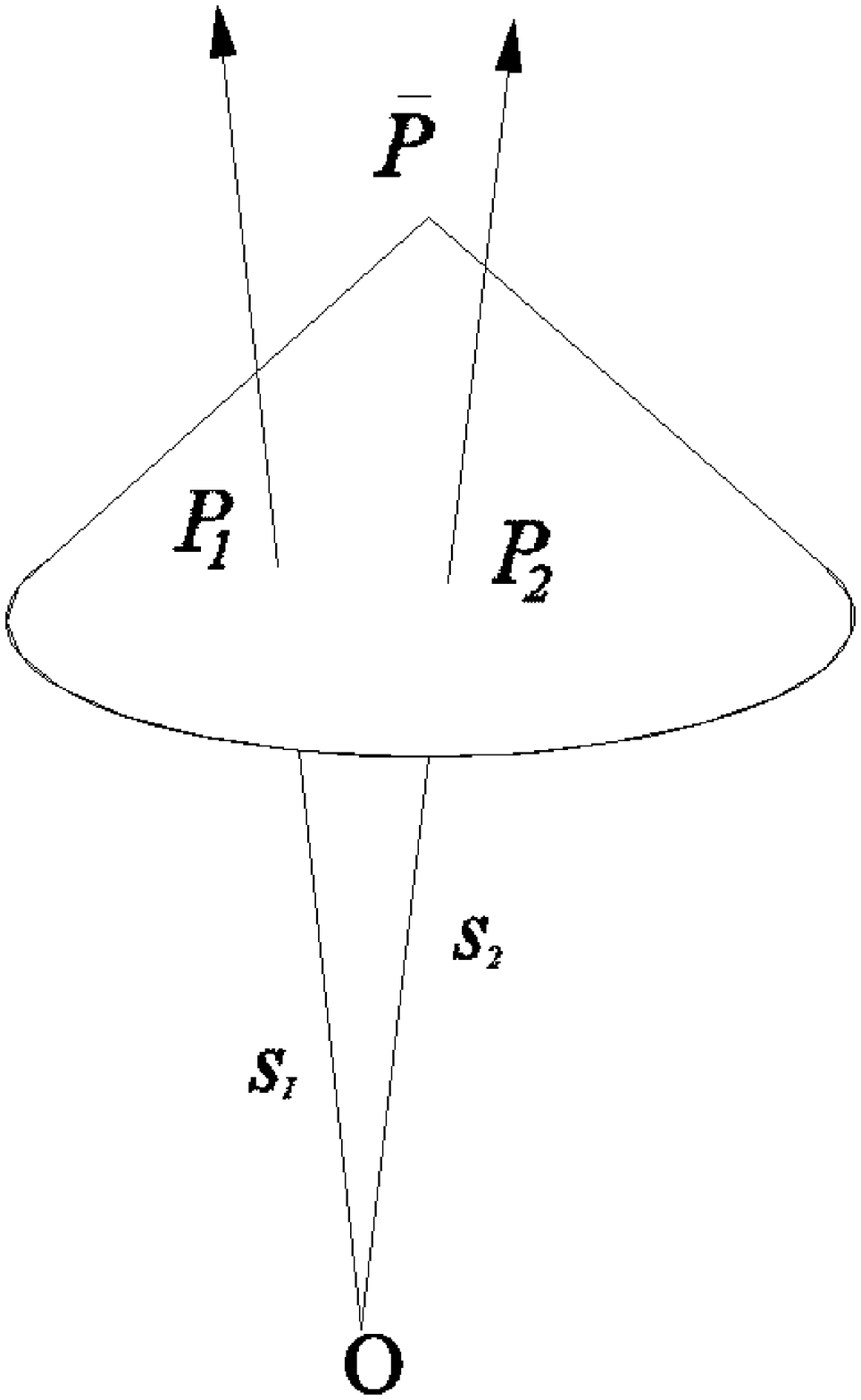}&
\includegraphics[scale=0.43]{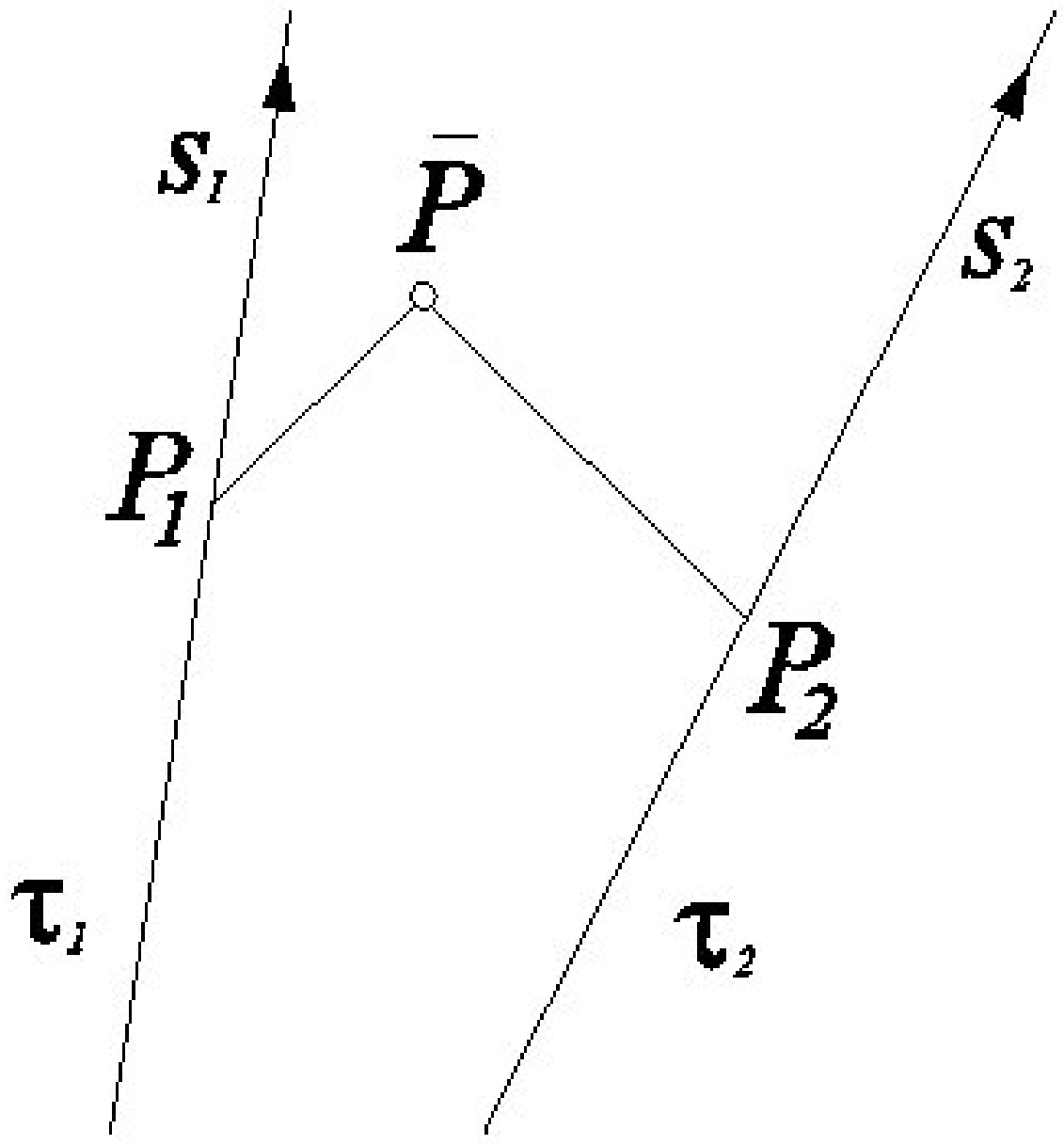}&\\[.2cm]
\mbox{(a)} & \mbox{(b)} 
\end{array}
$\\
\end{center}
\caption{
The satellite configuration  in flat spacetime is schematically shown in Fig. (a) by suppressing two spatial directions. The past light cone at a generic spacetime point ${\bar P}$ cuts the two satellite world lines $S_1$, $S_2$ at points $P_1$ and $P_2$, respectively.
Fig. (b) represents instead the construction of emission coordinates in a $1 + 1$-dimensional spacetime diagram.
}
\label{fig:1}
\end{figure}

\section{Fermi vs emission coordinates}

Consider a generic spacetime metric and introduce a Fermi coordinate system $(T,X,Y,Z)$ in some neighborhood of an accelerated world line with  (constant) acceleration ${\mathcal A}$; the spatial coordinates $X,Y,Z$ are associated with three Fermi-Walker dragged axes along the world line while $T$ measures proper time along the world line at the origin of the spatial coordinates.
Up to terms linear in the spatial coordinates, one has (see Eq. (6.18) of Ref. \cite{mtw})
\beq\fl
\label{fermimetric}
\rmd s^2=(\eta_{\alpha\beta}+2{\mathcal A}X\delta_\alpha^0\delta_\beta^0)\rmd X^\alpha\rmd X^\beta
=-(1-2{\mathcal A}X)\rmd T^2+\rmd X^2+\rmd Y^2+\rmd Z^2+O(2)\ , 
\eeq 
valid within a world tube region of radius $1/{\mathcal A}$ so that $|{\mathcal A} X| \ll 1$ is the condition for this approximation to be correct.

Let the gravitational field of the Earth be represented by the exterior Schwarzschild solution, whose metric written in standard coordinates $(t,r,\theta,\phi)$ is given by
\beq\fl\quad 
\label{Schwmetric}
\rmd  s^2 = -\left(1-\frac{2M}r\right)\rmd t^2 + \left(1-\frac{2M}r\right)^{-1} \rmd r^2 
+ r^2 (\rmd \theta^2 +\sin^2 \theta \rmd \phi^2)\ .
\eeq
The case of Schwarzschild spacetime cannot be treated explicitly, since the geodesics are not known in closed analytic form, so that an exact analytic expression for the world function cannot be obtained.

We are interested in casting the metric (\ref{Schwmetric}) in the form (\ref{fermimetric}) about the world line of an accelerated observer at rest at a fixed position on the equatorial plane. The map between Schwarzschild coordinates $(t,r,\theta,\phi)$ and  Fermi coordinates $(T,X,Y,Z)$ associated with an observer at rest at $r = r_0$, $\theta=\theta_0=\pi/2$, $\phi=\phi_0$ has been first derived by Leaute and Linet \cite{linet} up to second order corrections in the spatial Fermi coordinates \footnote{This result has been generalized later to the case of a static observer located at any point on the equatorial plane of the Kerr spacetime and to any uniformly rotating circular equatorial orbit by Bini, Geralico and Jantzen \cite{bgj}.}
\begin{eqnarray}\fl\quad
\label{schwtofermi}
t&=&t_0+\left(1-\frac{2M}{r_0}\right)^{-1/2}T+O(3)\ , \nonumber\\
\fl\quad
r&=&r_0+\left(1-\frac{2M}{r_0}\right)^{1/2}X+\frac12\left[\frac{M}{r_0^2}X^2+\frac{1}{r_0}\left(1-\frac{2M}{r_0}\right)(Y^2+Z^2)\right]+O(3)\ , \nonumber\\
\fl\quad
\theta&=&\frac{\pi}{2}+\frac{Y}{r_0}-\frac{1}{r_0^2}\left(1-\frac{2M}{r_0}\right)^{1/2}XY+O(3)\ , \nonumber\\
\fl\quad
\phi&=&\phi_0+\frac{Z}{r_0}-\frac{1}{r_0^2}\left(1-\frac{2M}{r_0}\right)^{1/2}XZ+O(3)\ ,
\end{eqnarray}
with the uniform acceleration ${\mathcal A}$ entering Eq. (\ref{fermimetric}) given by
\beq
{\mathcal A}=\frac{M}{r_0^2}\left(1-\frac{2M}{r_0}\right)^{-1/2}\ . 
\eeq
The above relations allow one to easily find out the explicit tranformation between Schwarzschild coordinates and emission coordinates, once the mapping expressing general Fermi coordinates and emission coordinates is constructed.
Our results thus apply in the case of the Schwarzschild and Kerr spacetimes as well.   
Furthermore, the transformation (\ref{schwtofermi}) can be easily improved by including higher order terms in the approximation \cite{bgj}, making the description even more accurate.

Let us consider now a set of four satellites whose world lines are chosen in such a way that they reduce to the flat spacetime configuration (\ref{UAflat}) in the limit of vanishing acceleration parameter ${\mathcal A}$. 
All relevant quantities are evaluated in the Appendix.
The four velocities are given by (see Eq. (\ref{Ufermi}))
\begin{eqnarray}\fl\quad
\label{Ufermi2}
{\mathcal U}_1&=&\cosh\alpha_1\partial_T+\sinh\alpha_1\partial_X+{\mathcal A}\tau^1\cosh\alpha_1\left(2\sinh\alpha_1\partial_T+\cosh\alpha_1\partial_X\right)\ , \nonumber\\
\fl\quad
{\mathcal U}_2&=&\cosh\alpha_2\partial_T+\sinh\alpha_2\partial_Y+{\mathcal A}\tau^2\cosh^2\alpha_2\partial_X\ , \nonumber\\
\fl\quad
{\mathcal U}_3&=&\cosh\alpha_3\partial_T+\sinh\alpha_3\partial_Z+{\mathcal A}\tau^3\cosh^2\alpha_3\partial_X\ , \nonumber\\
\fl\quad
{\mathcal U}_4&=&\partial_T+{\mathcal A}\tau^4\partial_X\  
\end{eqnarray}
to first order in ${\mathcal A}$.
By integrating these equations with respect to each proper time one easily gets the corresponding world lines
\begin{eqnarray}
X_1&=&S_1+{\mathcal A}(\tau^1)^2\cosh \alpha_1\left(\sinh \alpha_1\partial_T+\frac12\cosh \alpha_1\partial_X\right)\ , \nonumber\\
X_2&=&S_2+\frac12{\mathcal A}(\tau^2)^2\cosh^2 \alpha_2\partial_X\ , \nonumber\\
X_3&=&S_3+\frac12{\mathcal A}(\tau^3)^2\cosh^2 \alpha_3\partial_X\ , \nonumber\\
X_4&=&S_4+\frac12{\mathcal A}(\tau^4)^2\partial_X\ , 
\end{eqnarray}
where the zeroth order quantities $S_A$ are given by Eqs. (\ref{satflat2}) and (\ref{UAflat}).

Consider then a generic spacetime point $\bar P$ with coordinates ${\bar X}^\alpha$ and a photon emitted at the generic point $P_A$ with coordinates $X_A^\alpha$ along the world line of the $A$th satellite. 
The equations of null geodesics are listed in the Appendix (see Eq. (\ref{nullgeosfermi})).
Let $\bar \lambda$ be the value of the affine parameter which corresponds to the meeting point $\bar P$ according to 
\beq
\bar X^\alpha={\mathcal K}^\alpha\bar \lambda+X^\alpha_A\ ,  
\eeq
where the null vector ${\mathcal K}$ is given by Eq. (\ref{null_g}), implying that 
\beq
X^\alpha_A-\bar X^\alpha=-{\mathcal K}^\alpha\bar \lambda\ .  
\eeq
Taking the norm of both sides (with the metric components evaluated at $\bar P$) leads to
\beq\fl
\label{eqtauAfermi}
-(\tau^A)^2+\eta_{\alpha\beta}{\bar X}^\alpha{\bar X}^\beta-2\tau^A\eta_{\alpha\beta}U_A^\alpha {\bar X}^\beta+{\mathcal A}{\bar X}^0[{\bar X}^0{\bar X}^1+\tau^A(U_A^1{\bar X}^0-2U_A^0{\bar X}^1)]=0\ 
\eeq
to first order in ${\mathcal A}$, which generalizes the corresponding Eq. (\ref{eqtauA}) valid in the case of flat spacetime.\\ 
\noindent{\it $\bullet$ Emission vs Fermi coordinates}\\
Searching for solutions of the form 
\beq
\tau^A=\tau^A{}^{(0)}+{\mathcal A}\tau^A{}^{(1)}\ ,
\eeq
where $\tau^A{}^{(0)}$ is given by Eq. (\ref{soltauA}) with ${\bar W}\to {\bar X}$, we get
\beq
\tau^A{}^{(1)}=\frac{{\bar X}^0}{2}\frac{{\bar X}^0{\bar X}^1+\tau^A{}^{(0)}(U_A^1{\bar X}^0-2U_A^0{\bar X}^1)}{\tau^A{}^{(0)}+\eta_{\alpha\beta}U_A^\alpha {\bar X}^\beta}\ .
\eeq
The inverse of the tranformed metric at $\bar P$ follows easily
\beq
\label{invFmetnewcoord}
g^{AB}=g^{\alpha\beta}\frac{\partial \tau^A}{\partial {\bar X}^\alpha}\frac{\partial \tau^B}{\partial {\bar X}^\beta} 
=g^{AB}{}^{(0)}+{\mathcal A}g^{AB}{}^{(1)}\ 
\eeq
to first order in ${\mathcal A}$, where $g^{AB}{}^{(0)}$ is given by Eq. (\ref{gABinvflat}) with ${\bar W}\to {\bar X}$ and $\tau^A\to \tau^A{}^{(0)}$ and 
\beq\fl\quad
g^{AB}{}^{(1)}=\left[
\eta^{\alpha\beta}\left(\frac{\partial \tau^A{}^{(1)}}{\partial {\bar X}^\alpha}\frac{\partial \tau^B{}^{(0)}}{\partial {\bar X}^\beta}+
\frac{\partial \tau^A{}^{(0)}}{\partial {\bar X}^\alpha}\frac{\partial \tau^B{}^{(1)}}{\partial {\bar X}^\beta}\right)
-2{\bar X}^1\frac{\partial \tau^A{}^{(0)}}{\partial {\bar X}^0}\frac{\partial \tau^B{}^{(0)}}{\partial {\bar X}^0}
\right]\ .	
\eeq
It is easy to show by a direct calculation that the condition $g^{AA}=0$ is preserved.
The vanishing of the diagonal components of the contravariant spacetime metric once written in emission coordinates is actually a general property of real null dual frames \cite{rovelli,hehl}, which can be easily explained because the $\tau^A=\,$const hypersurfaces are tangent to the light cone by construction.
 
We are left expressing the spacetime coordinates in terms of the emission coordinates. 
The condition ensuring that the past light cone of $\bar P$ cuts the emitter world lines writes as 
\beq 
\label{lightconeacc}
\Omega(X_A,{\bar X})=0\ , \qquad X^0_A<{\bar X}^0\ ,
\eeq
where the world function is given by 
\begin{eqnarray}
\label{WFacc}\fl\quad
\Omega(X_A,{\bar X})&\simeq&\frac12\left[\eta_{\alpha\beta}+{\mathcal A}(X_A^1+{\bar X}^1)\delta^0_\alpha\delta^0_\beta \right](X^\alpha_A-{\bar X}^\alpha)(X^\beta_A-{\bar X}^\beta)\ \nonumber\\
\fl\quad
&=&\Omega_{\rm flat}(X_A,{\bar X})+\frac12{\mathcal A}(X_A^1+{\bar X}^1)(X^0_A-{\bar X}^0)^2\ 
\end{eqnarray}
to first order in the acceleration parameter ${\mathcal A}$ (see Eq. (\ref{WFaccdef})).\\ 
\noindent{\it $\bullet$ Fermi vs emission coordinates}\\
Eq. (\ref{lightconeacc}) gives rise to a set of four equations for the coordinates of $\bar P$. 
We look for solutions of such a system to first order in ${\mathcal A}$, i.e.,
\beq
{\bar X} \simeq {\bar W}+{\mathcal A}{\bar w}\ ,
\eeq
where the solution for ${\bar W}$ is given by Eqs. (\ref{solsistemaexpl})--(\ref{coeffsX0}).
The zeroth order equations (\ref{lightconeacc}) are obviously identically satisfied by the flat spacetime solution ${\bar W}$. 
The remaining set of equations for the first order quantities ${\bar w}^\alpha$ is given by
\begin{eqnarray}\fl\quad
0&=&{\bar w}^0\Lambda^1-{\bar w}^1+{\bar W}^0\left[\frac12{\bar W}^0-{\bar W}^1\Lambda^1\right]\ , \nonumber\\
\fl\quad
0&=&{\bar w}^0\Lambda^2-{\bar w}^2-{\bar W}^0{\bar W}^1\Lambda^2\ , \nonumber\\
\fl\quad
0&=&{\bar w}^0\Lambda^3-{\bar w}^3-{\bar W}^0{\bar W}^1\Lambda^3\ , \nonumber\\
\fl\quad
0&=&-2{\bar w}^0({\bar W}^0-\tau^4)+2\delta_{ij}{\bar w}^i{\bar W}^j+{\bar W}^0{\bar W}^1({\bar W}^0-2\tau^4)\ , 
\end{eqnarray}
where the quantities $\Lambda^i$ are given by Eq. (\ref{lambda_i}).
The corresponding solution turns out to be 
\beq
{\bar w}^0={\bar W}^0{\bar W}^1\ , \quad 
{\bar w}^1=\frac12({\bar W}^0)^2\ , \quad 
{\bar w}^2=0={\bar w}^3\ .
\eeq

The components $g_{AB}$ of the transformed metric turn out to be given by
\beq
\label{Fmetnewcoord}
g_{AB}=g_{\alpha\beta}\frac{\partial \bar X^\alpha}{\partial \tau^A}\frac{\partial \bar X^\beta}{\partial \tau^B}
=g_{AB}^{(0)}+O(2)\ ,  
\eeq
where the zeroth order metric $g_{AB}^{(0)}$ is given by Eq. (\ref{gABflat}) and the  first order metric $g_{AB}^{(1)}$ vanishes.
According to the terminology of Ref. \cite{rovelli} the components of the metric tensor are thus \lq\lq complete'' observables, since they are completely determined by any given set of four emission coordinates. 
The latter are instead \lq\lq partial'' observables, since they are directly measured quantities.
This is true if the spacetime metric, i.e., the gravity field, is exactly known (in any coordinate system), implying that the system of satellites would constitute an ideal positioning system. 
In practice the spacetime metric is not exactly known, and the satellite system itself has to be used to infer it, as discussed by Coll and collaborators \cite{tarantola}.
The constellation of satellites can thus serve for both positioning and measuring the spacetime metric by equipping the satellites with an accelerometer (measuring deviations from geodesic motion) and a gradiometer (measuring the strength of the gravitational field). 
By taking advantage from this additional information on the metric an optimization procedure has been developed in \cite{tarantola} to obtain the \lq\lq best observational gravitational field'' acting on the constellation.

\section{Concluding remarks}

We have considered the metric associated with a generic uniformly accelerated observer in any spacetime close enough to the world line of the observer himself. 
In particular, this can be taken as the metric describing the homogeneous gravitational field of the Earth.
We have expressed this metric in terms of the so called emission coordinates, i.e., the four proper times measured along the (timelike) geodesic world lines of four satellites, generalizing previous results valid for flat spacetime.
The present analysis has been mostly motivated by the relevance of using emission coordinates in the definition of a {\it relativistic} positioning system around the Earth. 

We have considered a particular (symmetric) configuration of satellite motion allowing certain simplifications of otherwise more involved formulas. 
However, our results can be easily generalized to arbitrary configurations of satellites corresponding to more realistic situations.
In fact, the resulting metric and all the possible usages associated with it can only be implemented  numerically in any case.
In this respect, we have thus provided an algorithm to construct an emission coordinate system at the disposal of a user in the close vicinity of the Earth's surface which also takes into account  the acceleration due to the Earth's gravity.

\section*{Acknowledgements}

This work has been supported by the Italian Gruppo Nazionale di Fisica Matematica of INDAM.
The authors are grateful to the anonimous referees for useful comments and suggestions.

\appendix 

\section{Geodesics of the Fermi metric}

We give here the general form of both timelike and null geodesics of the Fermi metric (\ref{fermimetric}) as well as the expression of the world function, the latter of which is not given in the literature.

The timelike geodesics can be written in the form 
\beq
X^\alpha\simeq S^\alpha+{\mathcal A}s^\alpha\ , 
\eeq 
to first order in the acceleration parameter ${\mathcal A}$, or explicitly
\begin{eqnarray}
\label{timegeosfermi}
T(\tau)&=&C\tau+T_0+{\mathcal A} C \tau(P^X\tau+X_0)=S^0+{\mathcal A}s^0\ , \nonumber\\
X(\tau)&=&P^X\tau+X_0+\frac12 {\mathcal A}C^2\tau^2=S^1+{\mathcal A}s^1\ , \nonumber\\
Y(\tau)&=&P^Y\tau+Y_0=S^2\ , \nonumber\\
Z(\tau)&=&P^Z\tau+Z_0=S^3\ ,
\end{eqnarray}
where 
\beq 
C=\left[1+(P^{X})^2+(P^{Y})^2+(P^{Z})^2\right]^{1/2}\ . 
\eeq
The zeroth order quantities correspond to those of Eq. (\ref{satflat}), i.e.,
\beq
S^\alpha=U^\alpha\tau+S_0^\alpha\ ,
\eeq
where now
\beq 
U=C\partial_T+P^X\partial_X+P^Y\partial_Y+P^Z\partial_Z\ .
\eeq 
Furthermore 
\beq 
s^\alpha=C\tau \left[(P^X\tau+X_0)\delta^\alpha_0 +\frac12C\tau\delta^\alpha_1\right]\ ,
\eeq
so that the unit vector tangent to the timelike geodesic world lines turns out to be
\beq
\label{time_g}
{\mathcal U}\simeq U+{\mathcal A}u\equiv U+{\mathcal A}C[(2P^X\tau+X_0)\partial_T+C\tau\partial_X]\ .
\eeq 

Similarly, the null geodesics are given by
\begin{eqnarray}
\label{nullgeosfermi}
T(\lambda)&=&E\lambda+T(0)+{\mathcal A}E\lambda[K^X\lambda+X(0)]\ , \nonumber\\
X(\lambda)&=&K^X\lambda+X(0)+\frac12 {\mathcal A}E^2\lambda^2\ , \nonumber\\
Y(\lambda)&=&K^Y\lambda+Y(0)\ , \nonumber\\
Z(\lambda)&=&K^Z\lambda+Z(0)\ ,
\end{eqnarray}
where $\lambda$ is an affine parameter and
\beq 
E=\pm \left[(K^{X})^2+(K^{Y})^2+(K^{Z})^2\right]^{1/2}\ . 
\eeq
The tangent vector to the photon path is thus given by
\beq\fl
\label{null_g}
{\mathcal K}\simeq K+{\mathcal A}k\equiv
E\partial_T+K^X\partial_X+K^Y\partial_Y+K^Z\partial_Z+
{\mathcal A}E[(2K^X\lambda+X_0)\partial_T+E\lambda\partial_X]\ .
\eeq 

Finally, with these explicit expressions of the geodesics and using the definition (\ref{wf_def}), it is easy to obtain the form of the world function
\begin{eqnarray}
\label{WFaccdef}\fl\quad
\Omega(X_A,X_B)&\simeq&\frac12\left[\eta_{\alpha\beta}+{\mathcal A}(X_A^1+X_B^1)\delta^0_\alpha\delta^0_\beta \right](X^\alpha_A-X_B^\alpha)(X^\beta_A-X_B^\beta)\ \nonumber\\
\fl\quad
&=&\Omega_{\rm flat}(X_A,X_B)+\frac12{\mathcal A}(X_A^1+X_B^1)(X^0_A-X^0_B)^2
\end{eqnarray}
to first order in the acceleration parameter ${\mathcal A}$, where $X_A$ and $X_B$ are two generic spacetime points connected by a geodesic path.

Eq. (\ref{time_g}) is quite general.
For our purposes we need to specify a set of four satellite world lines that reduce to the flat spacetime configuration (\ref{UAflat}) in the limit of vanishing acceleration parameter ${\mathcal A}$. 
Their four velocities correspond to the geodesics (\ref{timegeosfermi}), all starting from the origin at each value of their proper times set equal to zero (i.e., $S_{0A}^\alpha=0$):
\begin{eqnarray}\fl\quad
\label{Ufermi}
{\mathcal U}_1&=&\sqrt{1+(P^X)^2}\partial_T+P^X\partial_X+{\mathcal A}\sqrt{1+(P^X)^2}\tau^1\left(2P^X\partial_T+\sqrt{1+(P^X)^2}\partial_X\right)\ , \nonumber\\
\fl\quad
{\mathcal U}_2&=&\sqrt{1+(P^Y)^2}\partial_T+P^Y\partial_Y+{\mathcal A}[1+(P^Y)^2]\tau^2\partial_X\ , \nonumber\\
\fl\quad
{\mathcal U}_3&=&\sqrt{1+(P^Z)^2}\partial_T+P^Z\partial_Z+{\mathcal A}[1+(P^Z)^2]\tau^3\partial_X\ , \nonumber\\
\fl\quad
{\mathcal U}_4&=&\partial_T+{\mathcal A}\tau^4\partial_X\ , 
\end{eqnarray}
which give Eq. (\ref{Ufermi2}) after introducing the rapidity parametrization
\beq
P^X=\sinh\alpha_1\ , \qquad
P^Y=\sinh\alpha_2\ , \qquad
P^Z=\sinh\alpha_3\ .
\eeq

\end{document}